\documentclass[11pt,letterpaper]{article}

\IfFileExists{latexml.sty}{\usepackage{latexml}}{\newif\iflatexml\latexmlfalse}

\usepackage[T1]{fontenc}
\usepackage[utf8]{inputenc}
\iflatexml
\newcommand{\latopage}{}
\else
\usepackage{lmodern}                 %
\usepackage[lining]{lato}            %
\usepackage[sfdefault,lining]{carlito} %
\DeclareFontFamily{T1}{CrltW}{}
\DeclareFontShape{T1}{CrltW}{m}{n}{<-> Crlt-tlf-t1--base}{}
\DeclareFontShape{T1}{CrltW}{m}{it}{<-> Crlt-Italic-tlf-t1--base}{}
\DeclareFontShape{T1}{CrltW}{b}{n}{<-> Crlt-Bold-tlf-t1--base}{}
\DeclareFontShape{T1}{CrltW}{b}{it}{<-> Crlt-BoldItalic-tlf-t1--base}{}
\DeclareFontShape{T1}{CrltW}{m}{sl}{<-> ssub * CrltW/m/it}{}
\DeclareFontShape{T1}{CrltW}{b}{sl}{<-> ssub * CrltW/b/it}{}
\DeclareFontShape{T1}{CrltW}{bx}{n}{<-> ssub * CrltW/b/n}{}
\DeclareFontShape{T1}{CrltW}{bx}{it}{<-> ssub * CrltW/b/it}{}
\DeclareFontShape{T1}{CrltW}{bx}{sl}{<-> ssub * CrltW/b/it}{}
\DeclareFontFamily{TS1}{CrltW}{}
\DeclareFontShape{TS1}{CrltW}{m}{n}{<-> ssub * Crlt-TLF/m/n}{}
\DeclareFontShape{TS1}{CrltW}{m}{it}{<-> ssub * Crlt-TLF/m/it}{}
\DeclareFontShape{TS1}{CrltW}{b}{n}{<-> ssub * Crlt-TLF/b/n}{}
\DeclareFontShape{TS1}{CrltW}{b}{it}{<-> ssub * Crlt-TLF/b/it}{}
\DeclareFontShape{TS1}{CrltW}{bx}{n}{<-> ssub * Crlt-TLF/b/n}{}
\DeclareFontShape{TS1}{CrltW}{bx}{it}{<-> ssub * Crlt-TLF/b/it}{}

\newcommand{\latopage}{\fontfamily{\latofamily}\selectfont}
\usepackage{textcomp}
\usepackage[protrusion=false,expansion=false]{microtype} %
\DisableLigatures[f]{encoding = *, family = *}  %
\input{glyphtounicode}\pdfgentounicode=1
\pdfglyphtounicode{fi}{0066 0069}\pdfglyphtounicode{f_i}{0066 0069}\pdfglyphtounicode{fl}{0066 006C}\pdfglyphtounicode{f_l}{0066 006C}
\pdfglyphtounicode{ff}{0066 0066}\pdfglyphtounicode{f_f}{0066 0066}\pdfglyphtounicode{ffi}{0066 0066 0069}\pdfglyphtounicode{f_f_i}{0066 0066 0069}
\pdfglyphtounicode{uniFB00}{0066 0066}\pdfglyphtounicode{uniFB01}{0066 0069}\pdfglyphtounicode{uniFB02}{0066 006C}\pdfglyphtounicode{uniFB03}{0066 0066 0069}\pdfglyphtounicode{uniFB04}{0066 0066 006C}
\fi
\DeclareUnicodeCharacter{2014}{\textemdash\penalty\exhyphenpenalty}%

\usepackage[letterpaper,left=56.2bp,right=56.2bp,top=44.1bp,bottom=50.5bp,
            headheight=14bp,headsep=17.1bp,footskip=24bp]{geometry}
\usepackage[table]{xcolor}
\usepackage{graphicx}
\usepackage{array}
\usepackage{longtable}
\usepackage{enumitem}
\usepackage{titlesec}
\iflatexml\else\usepackage[titles]{tocloft}\fi
\usepackage{fancyhdr}
\usepackage{lastpage}
\usepackage{caption}
\usepackage{float}
\iflatexml\else\usepackage[absolute]{textpos}\fi
\usepackage[document]{ragged2e}
\iflatexml\else\usepackage[most]{tcolorbox}\fi
\usepackage[nocompress,nosort]{cite}  %
\usepackage{url}
\usepackage{etoolbox}
\usepackage[hidelinks]{hyperref}
\hypersetup{
  pdftitle={ChainIT Transaction Truth: Stablecoins, Web3 Wallets, and Multi-Rail Commerce. A Zero-Trust Control Plane for Verified Identity, Organizational Authority, Runtime Compliance, Secure Wallet Execution, and Immutable Settlement Evidence},
  pdfauthor={Eric Tacl, Matthew Koepp, Tim Hines, Russell Lessard, Jason Patrick, Jeremy Blackburn},
  pdfsubject={ChainIT Transaction Truth Architecture (CIT-TT-EWP-2.1)},
  pdfkeywords={stablecoins, Web3 wallets, multi-party computation, payment authorization, zero trust, verifiable credentials, transaction evidence, ChainIT}
}

\definecolor{body}{HTML}{1F2937}
\definecolor{cnavy}{HTML}{1B2A4A}
\definecolor{cblue}{HTML}{2B6CB0}
\definecolor{cgray}{HTML}{4A5568}
\definecolor{hgray}{HTML}{6B7280}
\definecolor{crule}{HTML}{D9DEE7}
\definecolor{rowalt}{HTML}{F2F6FA}
\definecolor{clink}{HTML}{1F6FB2}
\definecolor{bluebar}{HTML}{2B6CB0}\definecolor{bluefill}{HTML}{F2F6FA}
\definecolor{goldbar}{HTML}{B8891E}\definecolor{goldfill}{HTML}{FFF5D6}
\definecolor{greenbar}{HTML}{2D7D46}\definecolor{greenfill}{HTML}{EEF6EA}

\renewcommand{\normalsize}{\fontsize{11bp}{13.82bp}\selectfont}
\newcommand{\tabsize}{\fontsize{10bp}{12.2bp}\selectfont}
\newcommand{\refsize}{\fontsize{10.5bp}{13.2bp}\selectfont}
\AtBeginDocument{\normalsize\color{body}}
\iflatexml\else
\makeatletter\newcommand{\listafterheading}{\if@nobreak\vspace{0.85bp}\fi}\makeatother %
\setlist[itemize]{label=\textbullet,leftmargin=15.8bp,labelindent=5.7bp,labelsep=*,itemsep=0.6bp,parsep=0bp,topsep=1bp,partopsep=0bp,
  before={\fontsize{11bp}{13.4bp}\selectfont\listafterheading}}%
\fi
\newif\ifpanellast\newif\iftablelast
\iflatexml\else\AddToHook{para/begin}{\global\panellastfalse\global\tablelastfalse}\fi
\titleformat{\section}{\raggedright\hyphenpenalty=10000\bfseries\fontsize{16bp}{19.5bp}\selectfont\color{cblue}}{\rule{0bp}{15.8bp}\thesection.}{0.25em}{\rule{0bp}{15.8bp}}
\titleformat{\subsection}{\raggedright\hyphenpenalty=10000\bfseries\fontsize{13bp}{15.8bp}\selectfont\color{cblue}}{\rule{0bp}{12.8bp}\thesubsection}{0.25em}{\rule{0bp}{12.8bp}}
\titlespacing*{\section}{0bp}{\ifpanellast 19.1bp\else\iftablelast 12bp\else 5.6bp\fi\fi}{1.1bp}
\titlespacing*{\subsection}{0bp}{\ifpanellast 2.6bp\else\iftablelast 13.6bp\else 3.55bp\fi\fi}{-1.4bp}
\newcommand{\unnumbered}[1]{\section*{#1}\addcontentsline{toc}{section}{#1}}

\iflatexml\else

\renewcommand{\cftsecpagefont}{\mdseries}
\renewcommand{\cftdotsep}{0}
\renewcommand{\cftsecleader}{\mdseries\cftdotfill{\cftdotsep}}
\newsavebox{\tocnum}
\renewcommand{\cftsecpresnum}{\global\setbox\tocnum\hbox\bgroup}
\renewcommand{\cftsecaftersnum}{.\egroup}

\renewcommand{\cftsubsecpresnum}{\global\setbox\tocnum\hbox\bgroup}
\renewcommand{\cftsubsecaftersnum}{\egroup}

\renewcommand{\cftsecfillnum}[1]{{\cftsecleader}\nobreak\hbox{\kern1bp\cftsecpagefont #1}\cftsecafterpnum\par}
\renewcommand{\cftsubsecfillnum}[1]{{\cftsubsecleader}\nobreak\hbox{\kern1bp\cftsubsecpagefont #1}\cftsubsecafterpnum\par}
\makeatletter\renewcommand{\@tocrmarg}{14bp}\renewcommand{\@pnumwidth}{1.5em}\makeatother
\fi

\DeclareCaptionFont{navy}{\color{cnavy}}
\DeclareCaptionFont{capsize}{\fontsize{8.5bp}{10.4bp}\selectfont}
\iflatexml
\newcommand{\ttbox}[2]{\begin{quote}\textbf{#2}\end{quote}}
\newcommand{\ttformula}[2]{\begin{quote}\centering\textbf{#2}\end{quote}}
\else
\newcommand{\ttboxbase}[3]{%
  \par\kern-\prevdepth
  \begin{tcolorbox}[enhanced,breakable,frame hidden,interior hidden,boxrule=0bp,arc=0bp,boxsep=0bp,colupper=cnavy,
    left=9.1bp,right=6bp,top=7.7bp,bottom=5bp,before skip=\iftablelast 23.3bp\else 8.9bp\fi,after skip=22.3bp,lines before break=1,
    underlay={\fill[#1fill]([xshift=2.1bp]frame.south west) rectangle (frame.north east);
              \fill[#1bar](frame.south west) rectangle ([xshift=2.1bp]frame.north west);},
    before upper={#2\bfseries\fontsize{11bp}{13.45bp}\selectfont\clubpenalty=0\widowpenalty=0}]
  #3\rule[-3bp]{0bp}{0bp}%
  \end{tcolorbox}\global\panellasttrue}
\newcommand{\ttbox}[2]{\ttboxbase{#1}{\RaggedRight}{#2}}
\newcommand{\ttformula}[2]{\ttboxbase{#1}{\centering}{#2}}
\fi

\arrayrulecolor{crule}
\newlength{\padtop}\newlength{\padbot}
\newcommand{\cellpadtop}{\rule{0bp}{\padtop}}
\newcommand{\cellpadbot}{\rule[-\padbot]{0bp}{0bp}}
\makeatletter\iflatexml\else\AtBeginEnvironment{longtable}{\if@nobreak\setlength{\LTpre}{4.1bp}\else\setlength{\LTpre}{7.8bp}\ifpanellast\else\par\kern-\prevdepth\kern1.05bp\fi\fi}\fi\makeatother %
\iflatexml\else\AfterEndEnvironment{longtable}{\global\tablelasttrue}\fi
\newcolumntype{P}[1]{>{\RaggedRight\arraybackslash\tabsize\cellpadtop}m{#1}<{\cellpadbot}}
\newcommand{\thc}[1]{\rule[-9.8bp]{0bp}{27.1bp}\textcolor{white}{\bfseries\fontsize{11bp}{13.5bp}\selectfont #1}}
\newcommand{\thcc}[1]{\color{white}\bfseries\fontsize{11bp}{13.5bp}\selectfont\rule{0bp}{13.9bp}#1\rule[-6.5bp]{0bp}{0bp}}%
\newcommand{\headrow}{\rowcolor{cnavy}}

\iflatexml
\newcommand{\refurlx}[2]{\href{#1}{#1}}
\newcommand{\refurlbreak}[2]{}
\else
\newcommand{\refurlbreak}[2]{\par\nobreak\prevdepth=\dimexpr\prevdepth+1.6bp\relax\refurlx{#1}{#2}\par\prevdepth=\dimexpr\prevdepth+0.3bp\relax}
\makeatletter
\newcommand{\refurlx}[2]{{\fontsize{9bp}{12.8bp}\selectfont\color{clink}\href{#1}{\refseg@#2-\@nil}}}
\def\refseg@#1-#2\@nil{\ifx\relax#2\relax\ifx\relax#1\relax\else\refword{#1}\fi\else\refword{#1-}\penalty-1 \refseg@#2\@nil\fi}
\newcommand{\refword}[1]{\setbox0\hbox{#1}\rlap{\rule[-1.55bp]{\wd0}{0.6bp}}\box0}
\makeatother
\fi
\newcommand{\refurl}[1]{\refurlx{#1}{#1}}
\iflatexml\else
\makeatletter
\renewenvironment{thebibliography}[1]
  {\list{[\@arabic\c@enumiv]}{\leftmargin12.9bp\labelwidth0bp\labelsep0bp\itemindent-12.9bp\listparindent0bp
     \itemsep5.5bp\parsep0bp\topsep0bp\partopsep0bp\usecounter{enumiv}}\RaggedRight}
  {\endlist}
\makeatother
\fi

\title{ChainIT Transaction Truth: Stablecoins, Web3 Wallets, and Multi-Rail Commerce\\[2pt]\large A Zero-Trust Control Plane for Verified Identity, Organizational Authority, Runtime Compliance, Secure Wallet Execution, and Immutable Settlement Evidence}
\newcommand{\affil}{ChainIT, Utah, United States (\url{https://chainit.com})}%
\author{Eric Tacl\thanks{\affil} \and Matthew Koepp\thanks{\affil} \and Tim Hines\thanks{\affil} \and Russell Lessard\thanks{\affil} \and Jason Patrick\thanks{\affil} \and Jeremy Blackburn\thanks{\affil}}
\date{September 17, 2026}

\iflatexml\else
\renewenvironment{abstract}{\section*{Abstract}}{\par}
\fi

\begin{document}

\iflatexml
\maketitle
\else
\thispagestyle{empty}
\setlength{\TPHorizModule}{1bp}\setlength{\TPVertModule}{1bp}
\newcommand{\coverline}[3]{%
  \begin{textblock*}{499.6bp}(56.2bp,#1bp)\centering #2 #3\par\end{textblock*}}
\coverline{46.9}{\bfseries\fontsize{30bp}{36bp}\selectfont\color{cnavy}}{ChainIT Transaction Truth}
\coverline{87.9}{\bfseries\fontsize{16bp}{19.5bp}\selectfont\color{cblue}}{Stablecoins, Web3 Wallets, and Multi-Rail Commerce}
\coverline{112.0}{\fontsize{11bp}{13.5bp}\selectfont\color{cgray}}{A Zero-Trust Control Plane for Verified Identity, Organizational Authority, Runtime Compliance, Secure Wallet\\Execution, and Immutable Settlement Evidence}
\coverline{209.1}{\bfseries\fontsize{16bp}{19.5bp}\selectfont\color{cgray}}{ChainIT}
\coverline{230.6}{\bfseries\fontsize{16bp}{19.5bp}\selectfont\color{cblue}}{\url{https://chainit.com}}
\coverline{310.1}{\bfseries\fontsize{11bp}{13.5bp}\selectfont\color{cnavy}}{CIT-TT-EWP-2.1}
\coverline{323.8}{\itshape\fontsize{9.5bp}{11.5bp}\selectfont\color{cgray}}{Executive Core Edition}
\coverline{444.0}{\fontsize{10.5bp}{12.5bp}\selectfont\color{hgray}}{September 17, 2026}
\coverline{461.9}{\itshape\fontsize{8bp}{9.7bp}\selectfont\color{hgray}}{Companion documents: ChainIT \href{https://chainit.com/insights/chainit-launches-provable-authority-as-u-s-generative-ai-enabled-fraud-losses-are-projected-to-reach-40-billion-by-2027/}{Provable Authority} (CIT-PA-WP-7.0); ChainIT \href{https://chainit.com/white-papers/}{Provable Compliance} (CIT-PC-WP-4.0); and the ChainIT Transaction Truth\\Technical and Compliance Companion (CIT-TT-TCC-2.1).}
\coverline{535.4}{\bfseries\fontsize{11bp}{13.5bp}\selectfont\color{body}}{Authors}
\coverline{553.5}{\fontsize{11bp}{13.45bp}\selectfont\color{body}}{\href{https://chainit.com/about-us/}{Dr. Eric Tacl}, Executive Vice President- Verified Payments and Commerce\\\href{https://chainit.com/about-us/}{Matthew Koepp}, Chief Technology Officer\\\href{https://chainit.com/about-us/}{Tim Hines}, Chief Product Officer\\\href{https://chainit.com/about-us/}{Russell Lessard}, Chief Compliance Officer\\\href{https://chainit.com/about-us/}{Jason Patrick}, Chief Commercial Officer\\\href{https://chainit.com/about-us/}{Jeremy Blackburn}, Chief Executive Officer}
\coverline{662.1}{\fontsize{7.5bp}{9.1bp}\selectfont\color{hgray}}{This paper describes the ChainIT \href{https://chainit.com/white-papers/}{Transaction Truth} Architecture at an executive and governance level. Detailed protocol, wallet, MPC, compliance, threat-model,\\and implementation-status material remains in the companion technical paper. This document does not constitute legal, regulatory, banking, custody, accounting,\\security-certification, or patentability advice.}
\coverline{695.0}{\bfseries\fontsize{8.5bp}{10.2bp}\selectfont\color{body}}{\textcopyright{} 2026, Scientia Potentia Est II, LLC. All rights reserved.}
\coverline{708.5}{\fontsize{7.5bp}{9.1bp}\selectfont\color{hgray}}{ChainIT is a registered trademark of Scientia Potentia Est II, LLC. Distributed by arXiv under the arXiv.org perpetual, non-exclusive license.}
\null\clearpage
\fi
\color{body}

\begin{abstract}
Stablecoins can move value continuously across programmable networks, and Web3 wallets can authorize those transfers with cryptographic signatures, but neither capability, standing alone, establishes a complete commercial transaction. A blockchain may prove that a transaction satisfied protocol rules, and a wallet may prove that sufficient signing authority existed; neither proves who initiated the transaction, for which legal entity, under what current authority, against which current compliance evidence, on what exact instruction, or with what settlement and accounting outcome. We describe the ChainIT \href{https://chainit.com/white-papers/}{Transaction Truth} Architecture, a zero-trust control, execution, and evidence architecture that places verified individual and organizational identity, the organizational authority record and authority policy of the ChainIT \href{https://chainit.com/insights/chainit-launches-provable-authority-as-u-s-generative-ai-enabled-fraud-losses-are-projected-to-reach-40-billion-by-2027/}{Authority Protocol}, transaction-specific runtime compliance decisions, exact-digest approval, atomic capacity reservation, and a single-use Execution Authorization Credential in front of wallet execution, and that records settlement, reconciliation, and lifecycle evidence after the transaction. The architecture separates five boundaries: control of a key is not identity, identity is not organizational authority, authority is not compliance, compliance is not execution authority, and network confirmation is not complete commercial settlement. We describe wallet and custody profiles, including a multi-party computation (MPC) profile in which custody security is kept independent of business authority, stablecoin and Web3 evidence domains, layered finality, the U.S. stablecoin regulatory context as of September 2026, and a phased deployment in which pilot success requires that the control invariants hold under attack and failure.
\end{abstract}


{\fontsize{10bp}{12.2bp}\selectfont\color{cgray}This paper was first released as a ChainIT white paper (CIT-TT-EWP-2.1). The companion papers, \emph{Provable Authority}~\cite{tt1} and \emph{Provable Compliance}~\cite{tt2}, and other ChainIT white papers are listed at \url{https://chainit.com/white-papers/}.\par}
\color{body}

\ttbox{blue}{Document precedence: CIT-PA-WP-7.0 controls shared \href{https://chainit.com/insights/chainit-launches-provable-authority-as-u-s-generative-ai-enabled-fraud-losses-are-projected-to-reach-40-billion-by-2027/}{Authority Protocol} definitions and flows. CIT-PC-WP-4.0 controls shared \href{https://chainit.com/white-papers/}{Provable Compliance} definitions and flows. The ChainIT Transaction Truth Technical and Compliance Companion, CIT-TT-TCC-2.1, controls the detailed wallet, MPC, execution, threat-model, schema, and architecture-status material. This executive paper does not modify those documents.}

Architecture status is stated by class: current platform foundation, published protocol architecture, development or MVP, proposed execution profile, and external dependency. This edition does not represent that every described component is generally available in every environment.
\clearpage

\iflatexml\else
\tableofcontents
\clearpage
\fi

\section*{\rule{0bp}{22.3bp}Document Precedence and Key Terms}\addcontentsline{toc}{section}{Document Precedence and Key Terms}%
Authority definitions control. ChainIT \href{https://chainit.com/insights/chainit-launches-provable-authority-as-u-s-generative-ai-enabled-fraud-losses-are-projected-to-reach-40-billion-by-2027/}{Provable Authority}, CIT-PA-WP-7.0, controls every shared authority term, subject classification, credential, decision artifact, and pre-execution process used in this paper. The Authority Protocol applies to individual, organizational, group, scheduled-workflow, and registered-agent activity across fiat, card, instant-payment, stablecoin, and on-chain transactions. \cite{tt1}

Compliance definitions control. ChainIT \href{https://chainit.com/white-papers/}{Provable Compliance}, CIT-PC-WP-4.0, controls every shared compliance artifact, evidence class, policy profile, disposition, freshness rule, and disclosure boundary used in this paper. A Compliance Decision \href{https://chainit.com/product/data-tokenization/}{Validated Data Tokens} (VDT) supplies one signed, time-bounded predicate to the \href{https://chainit.com/insights/chainit-launches-provable-authority-as-u-s-generative-ai-enabled-fraud-losses-are-projected-to-reach-40-billion-by-2027/}{Authority Protocol}. It does not create authority and does not independently authorize payment. \cite{tt2}

Naming hierarchy and technical companion. ChainIT \href{https://chainit.com/white-papers/}{Transaction Truth} is the umbrella name; the ChainIT Transaction Truth Architecture is the formal technical reference; and the Transaction Truth control plane is the functional pre-execution layer. CIT-TT-TCC-2.1 controls wallet profiles, MPC, exact-transaction mechanics, asset and chain controls, the threat model, pilot criteria, canonical object schemas, the ChainIT Wallet Execution Profile, and implementation status. \cite{tt3,tt4}

\rowcolors{2}{white}{white}
\begin{longtable}{|@{\hspace{4bp}}P{207.8bp}|P{275.0bp}@{\hspace{4bp}}|}
\hline
\headrow\thc{Term} & \thc{Executive meaning}\\\hline
\endfirsthead
\hline
\headrow\thc{Term} & \thc{Executive meaning}\\\hline
\endhead
\textbf{Allow} & The final compliance disposition assigned by the organization\textquotesingle{}s Workflow Engine when every predicate required by the applicable Compliance Policy Profile has returned a current, reason-coded true result. Allow is one of the enumerated dispositions of the Compliance Protocol, alongside Allow with Controls, Step-Up, Review, Hold, Reject, Prohibit, and Stale/Unknown. Only a final Allow sets Compliance Decision Permits to true in the \href{https://chainit.com/insights/chainit-launches-provable-authority-as-u-s-generative-ai-enabled-fraud-losses-are-projected-to-reach-40-billion-by-2027/}{Authority Protocol}. Allow permits compliance-side continuation for one reliance event only; it does not establish identity, confer authority, cure an altered digest or exceeded limit, or authorize execution. Any disposition other than Allow, including missing, stale, ambiguous, or contradictory evidence, is a controlled non-Allow state and stops the transaction from proceeding on the compliance side.\\\hline
\textbf{Authority Resolution \href{https://www.pactvera.com/}{Pactvera} (ARP)} & The canonical, versioned organizational authority record: who may act, in what role, for what purpose, within what limits, during what period, under what quorum or separation-of-duties requirements, and subject to what amendment or revocation?\\\hline
\textbf{Authority Policy Profile (APP)} & The versioned authority-side rule set selected for a reliance event, varying by organization, role, subject, counterparty, transaction, value, rail, jurisdiction, risk tier, and purpose.\\\hline
\textbf{Authority Protocol} & The complete pre-execution process by which a verified subject\textquotesingle{}s present authority to perform one exact action is resolved from the Authority Resolution \href{https://www.pactvera.com/}{Pactvera} under the applicable Authority Policy Profile, combined with the Compliance Decision, required approvals, and capacity reservation, and expressed as an Execution Authorization Credential. The \href{https://chainit.com/insights/chainit-launches-provable-authority-as-u-s-generative-ai-enabled-fraud-losses-are-projected-to-reach-40-billion-by-2027/}{Authority Protocol} determines who may act; it is distinct from the Authority Policy Profile, which is the versioned rule set the Protocol applies at a given reliance event.\\\hline
\textbf{Compliance Policy Profile (CPP)} & The versioned compliance-side rule set selecting required evidence, freshness bounds, predicates, review conditions, disclosure rules, and transaction-specific controls.\\\hline
\textbf{Compliance Protocol} & The runtime evaluation of current evidence against the applicable Compliance Policy Profile described as “\href{https://chainit.com/white-papers/}{Provable Compliance},” produces a transaction-specific, time-bounded decision from current evidence and policy.\\\hline
\textbf{Business Rules Engine (BRE)} & The deterministic predicate-evaluation engine. It returns explicit, reason-coded true or false results; it does not assign the organization\textquotesingle{}s final disposition.\\\hline
\textbf{Compliance Decision VDT} & A signed, time-bounded, purpose-specific record of the organization-controlled compliance disposition for one reliance event.\\\hline
\textbf{Canonical Transaction Digest} & A deterministic hash of the exact transaction object under a fixed encoding rule so any material change produces a different digest.\\\hline
\textbf{Evidence Assertion VDT} & A signed, time-stamped record of a single compliance fact, such as \textquotedbl{}this person passed identity verification on this date\textquotedbl{} or \textquotedbl{}this entity was screened against the sanctions list with no match.\textquotedbl{} Each one records where the fact came from, when it was observed, how long it stays valid, what it was gathered for, and how confident the source was. These records are the evidence the compliance process weighs; the decision itself is recorded separately as a Compliance Decision \href{https://chainit.com/product/data-tokenization/}{VDT}.\\\hline
\textbf{Execution Authorization Credential (EAC)} & A short-lived, audience-restricted, single-use credential authorizing one exact transaction, bound to the digest, reservation, subject or session, status, and expiration.\\\hline
\textbf{GENIUS Act} & Guiding and Establishing National Innovation for U.S. Stablecoins Act.\\\hline
\textbf{Multi-Party Computation (MPC)} & A cryptographic custody technique in which a private key is never assembled in one place; instead, the key is split into shares held by separate parties or systems, and a defined threshold of those shares must cooperate to produce a valid signature. MPC reduces single-key compromise risk and allows signer distribution across infrastructure, custodians, or devices. In the \href{https://chainit.com/white-papers/}{Transaction Truth} Architecture, MPC operates only within the wallet and custody plane: satisfying the cryptographic signing threshold is independent of, and never substitutes for, verified identity, organizational authority under the ARP, a current compliance decision, or an exact-digest match.\\\hline
\textbf{Wallet Execution Profile (WEP)} & The versioned wallet-specific control profile governing custody model, signer requirements, asset and chain scope, recovery, limits, destination policy, and evidence.\\\hline
\textbf{Pactvera / \href{https://chainit.com/product/data-tokenization/}{VDT} / Valitorum} & \href{https://www.pactvera.com/}{Pactvera} governs transaction logic; VDTs record validated data, decisions, and lifecycle events; \href{https://chainit.com/}{Valitorum} seals the completed or otherwise terminal evidence chain without rewriting prior records.\\\hline
\textbf{Reliance Event} & The moment a party relies on identity, authority, evidence, policy, or transaction state. Freshness and present eligibility are evaluated at that moment, not inherited indefinitely from onboarding.\\\hline
\textbf{Session Credential} & The token or cookie a system issues after a user logs in so that it can recognize the user across subsequent requests without re-authentication. It proves only that an earlier login succeeded. In the \href{https://chainit.com/white-papers/}{Transaction Truth} Architecture, a session credential never authorizes a payment; authorization requires a separate Execution Authorization Credential bound to the exact transaction.\\\hline
\textbf{Transaction Truth Architecture} & The zero-trust control, execution, and evidence architecture that binds verified identity, current authority, transaction-specific compliance, the exact transaction digest, qualified wallet execution, settlement, reconciliation, and sealed evidence across four independently evaluated planes, so that value moves only on a commercially authorized transaction and the result is provable.\\\hline
\end{longtable}
\ttbox{gold}{Five boundaries remain separate: 1) control of a key is not identity; 2) identity is not organizational authority; 3) authority is not compliance; 4) compliance is not execution authority; and 5) network confirmation is not complete commercial settlement.}
\clearpage %
\unnumbered{Executive Summary}
Stablecoins can transfer value continuously across programmable networks. Web3 wallets can authorize those transfers through cryptographic signatures. Neither capability, standing alone, establishes a complete commercial transaction.

A blockchain may prove that a transaction satisfied certain rules. A wallet may prove that sufficient cryptographic authority existed to sign it. Those facts do not independently prove who initiated the transaction, for which legal entity, under what current authority, against which compliance evidence, subject to what business terms, or whether the merchant ultimately received and reconciled the intended economic value.

\ttbox{green}{Stablecoins and tokenized deposits are settlement instruments. Web3 wallets are cryptographic execution endpoints. ChainIT supplies the verified identity, organizational authority, runtime compliance, exact-transaction control, secure execution, multi-rail orchestration, reconciliation, and immutable evidence required to make them commercially usable.}
ChainIT addresses the gap through the ChainIT \href{https://chainit.com/white-papers/}{Transaction Truth} Architecture: a zero-trust control, execution, and evidence architecture connecting \href{https://chainit.com/product/identity-verification/}{ChainIT ID}, \href{https://chainit.com/product/business-verification/}{ChainIT Org ID}, the Authority Resolution \href{https://www.pactvera.com/}{Pactvera}, Provable Compliance, exact-transaction decisioning, secure wallet execution, \href{https://chainit.com/product/digital-pay/}{ChainIT Pay}, Pactvera, lifecycle \href{https://chainit.com/product/data-tokenization/}{VDTs}, ERP integration, and \href{https://chainit.com/}{Valitorum}.

\vspace{-3bp}\section*{Three controlling propositions}\vspace{-1.7bp}%
First, the wallet is downstream of authority and compliance. \href{https://chainit.com/product/identity-verification/}{ChainIT ID} and \href{https://chainit.com/product/business-verification/}{ChainIT Org ID} validate the relevant person and entity. The ARP and APP resolve whether the individual, group, scheduled workflow, or registered agent may perform the proposed action. The Compliance Protocol produces a transaction-specific, time-bounded decision from current evidence and policy. Only the complete \href{https://chainit.com/insights/chainit-launches-provable-authority-as-u-s-generative-ai-enabled-fraud-losses-are-projected-to-reach-40-billion-by-2027/}{Authority Protocol} can authorize execution. \cite{tt1,tt2}

Second, approval is bound to the exact instruction eventually signed. The payer, source, payee, destination, amount, asset, chain, token contract, fees, validity window, nonce, and other material parameters are canonicalized and hashed. Human or quorum approval applies to that digest. Capacity is reserved. A one-time EAC is issued. The wallet boundary reconstructs the digest and rejects any mismatch, replay, wrong audience, stale decision, or expired authorization.

Third, custody security and governance security are independent controls. MPC and threshold cryptography can reduce single-key compromise risk. Confidential-computing environments can isolate sensitive signing operations and attest approved code. Neither establishes legal or organizational authority. Conversely, a valid corporate approval does not create a blockchain signature. A controlled transaction requires both governance approval and qualified cryptographic execution.

\ttformula{green}{Commercially Authorized Transaction = Verified Subject AND Current Authority AND Compliance Decision Permits AND Exact Digest Match AND Required Approvals AND Capacity Reserved AND Qualified Wallet Execution}
The resulting model is issuer-agnostic, chain-agnostic, wallet-agnostic, bank-compatible, tokenized-deposit compatible, and multi-rail. It is not indiscriminate. Each asset, issuer, token contract, chain, bridge, wallet profile, custodian, bank, liquidity provider, and settlement path remains subject to qualification and current policy.

ChainIT therefore does not need to predict which stablecoin, blockchain, bank token, custodian, wallet, or instant-payment system will dominate. ChainIT can monetize the verified control, execution, and evidence layer across every qualified rail.

\section{Market Transition: New Rails, Same Commercial Obligations}
\subsection{A multi-money operating environment}
Enterprise payments are moving toward a hybrid environment in which cards, Automated Clearing House (ACH), wires, instant bank payments, stablecoins, tokenized deposits, internal ledgers, and future tokenized-money instruments coexist. Different instruments will remain useful for different purposes. Consumer credit, rewards, dispute rights, bank settlement, cross-border availability, programmability, operating hours, and cost do not collapse into one universal rail.

Stablecoins can improve speed, programmability, cross-border reach, and continuous availability. Tokenized deposits can provide bank-native digital settlement. Neither eliminates the commercial duties attached to identity, authorization, sanctions, contractual performance, custody, accounting, tax, audit, or dispute handling.

\subsection{Stablecoin, blockchain, and wallet are distinct components}
Market shorthand often calls stablecoin a payment rail. It is more precise to separate three components. The stablecoin is the settlement asset: a liability of its issuer that represents the value being transferred. The blockchain is the rail: the shared ledger that records and moves that value from one holder to another. The wallet is the control point: the software or service that holds the keys and signs the instruction to move it. These three layers may be supplied by different organizations, and each carries different legal, operational, and risk characteristics.

\rowcolors{2}{white}{rowalt}
\begin{longtable}{|@{\hspace{4bp}}P{143.1bp}|P{143.1bp}|P{188.1bp}@{\hspace{4bp}}|}
\hline
\headrow\thc{\raisebox{1.6bp}{Component}} & \thc{\raisebox{1.6bp}{Primary function}} & \thcc{Unresolved without \href{https://chainit.com/white-papers/}{Transaction Truth} controls}\\\hline
\endfirsthead
\hline
\headrow\thc{\raisebox{1.6bp}{Component}} & \thc{\raisebox{1.6bp}{Primary function}} & \thcc{Unresolved without \href{https://chainit.com/white-papers/}{Transaction Truth} controls}\\\hline
\endhead
\textbf{Stablecoin issuer} & Issues and redeems the asset; administers reserves, token contracts, and issuer controls. & Does not inherently prove the payer\textquotesingle{}s legal identity, corporate authority, commercial purpose, performance, or ERP result.\\\hline
\textbf{Blockchain or network} & Orders transactions and updates ledger state according to protocol rules. & Does not inherently know the legal parties, governing contract, organizational approvals, or off-chain outcome.\\\hline
\textbf{Wallet or custodian} & Controls credentials, keys, accounts, or signing policy. & Does not by itself establish current authority, transaction-specific compliance, or commercial settlement.\\\hline
\textbf{Bank or payment rail} & Moves fiat or maintains a deposit relationship under its own rules. & Does not necessarily preserve portable, cross-rail evidence of the complete transaction.\\\hline
\textbf{ChainIT \href{https://chainit.com/white-papers/}{Transaction Truth} Architecture} & Binds verified parties, authority, current evidence, exact transaction, wallet execution, settlement, reconciliation, and evidence. & Does not replace the issuer, network, bank, custodian, or regulated institution; it controls and connects qualified participants.\\\hline
\end{longtable}
\subsection{Stablecoins and tokenized deposits remain legally and economically distinct}
A payment stablecoin and a tokenized bank deposit may support similar user experiences, but they are not the same instrument. A stablecoin is generally an issuer obligation represented by a token and supported by eligible reserve assets. A tokenized deposit represents a deposit relationship with a regulated depository institution. ChainIT preserves that distinction in product disclosures, asset registries, custody treatment, accounting records, settlement instructions, risk controls, and customer interfaces. \cite{tt11}

The ChainIT opportunity is therefore not to replace every payment instrument. It is to maintain one verified transaction-control model while selecting the qualified settlement instrument best suited to the customer, merchant, jurisdiction, value, risk, and operating condition.

\section{The Control Gap: A Wallet Signature Is Not a Complete Authorization}
\subsection{Cryptographic validity answers a narrow question}
A blockchain checks that a transaction is technically correct: the signature is valid, the sender has the funds, the fees are paid, and any smart-contract rules are met. A multi-party wallet can produce that signature by having several key holders approve together. A smart account can approve it through rules written into the account itself. All of these answer the same narrow question: should the network accept this instruction? None of them answers the question a business actually cares about: should this instruction have been created in the first place, by this person, for this company, under this authority, and on today\textquotesingle{}s compliance evidence?

Private-key or account control can be exercised by the legitimate owner, an authorized employee, an employee acting outside authority, a former employee, a compromised device, an infrastructure operator, a malicious insider, a scheduled workflow, or an AI agent acting outside delegated purpose. The chain usually cannot distinguish among these cases.

\subsection{Address identity is not legal identity}
A blockchain address is a protocol identifier. It may be pseudonymous, rotated, shared, derived, administered by a custodian, or controlled by smart-contract logic. It is not inherently a legal name, verified individual, registered organization, beneficial owner, officer, fiduciary, licensee, or authorized representative.

\href{https://chainit.com/product/identity-verification/}{ChainIT ID} supplies the verified individual anchor. \href{https://chainit.com/product/business-verification/}{ChainIT Org ID} supplies the legal-entity anchor. Wallet-binding evidence links a wallet or account to the relevant subject under a defined assurance method. The evidence must also state what was not proven: a successful wallet challenge can demonstrate present control of a credential without proving source of funds, beneficial ownership of every asset, or authority to use organizational property.

\subsection{Authentication is not authority}
A person can authenticate correctly and still lack authority for the proposed action. A treasury analyst may prepare but not release a payment. A controller may approve ACH but not an irreversible stablecoin transfer. An employee may use an existing vendor but not change the destination address. An officer may be over an amount limit. A delegate may remain technically enrolled after the principal\textquotesingle{}s authority changes.

At the moment of the transaction, the ARP and APP are checked together: the ARP confirms that this person or agent currently holds the role and authority in question, and the APP confirms that the rules applicable to this transaction permit them to use it. Authority exists only where both agree. It is not inferred from the fact that someone was onboarded in the past or holds a wallet credential. \cite{tt1}

\subsection{Compliance is not a permanent badge}
A person, organization, wallet, issuer, asset, chain, jurisdiction, or counterparty is not universally and permanently \textquotedbl{}compliant.\textquotedbl{} Evidence changes and is purpose-specific. A sanctions result, entity status, customer classification, wallet-risk observation, license, issuer status, or jurisdictional permission may be sufficient for one transaction and insufficient for another.

\href{https://chainit.com/white-papers/}{Provable Compliance} therefore evaluates current Evidence Assertion \href{https://chainit.com/product/data-tokenization/}{VDTs} against the applicable CPP at the reliance event. The organization\textquotesingle{}s Workflow Engine composes BRE predicate results into a controlled disposition. A final Allow permits compliance-side continuation only; it does not cure absent authority, an altered digest, an exceeded limit, or an invalid wallet instruction. \cite{tt2}

\subsection{Network confirmation is not complete commercial settlement}
An on-chain transfer may be confirmed while the commercial transaction remains incomplete. The merchant may have requested fiat settlement, conversion may fail, a payout may not post, a contractual condition may remain open, or the ERP record may not reconcile. The ChainIT \href{https://chainit.com/white-papers/}{Transaction Truth} Architecture separates network acceptance from economic, commercial, accounting, and evidentiary completion.

\ttbox{gold}{Zero trust in this context means no person, device, wallet, address, custodian, asset, chain, policy result, or prior approval receives implicit authority. Each consequential transaction is re-evaluated against current identity, authority, evidence, policy, exact instruction, capacity, wallet profile, and outcome.}
\section{ChainIT Strategic Position}
\subsection{Control the transaction, not the monetary instrument}
Stablecoin issuance is regulated, reserve-intensive, distribution-driven, and dependent on redemption confidence, liquidity, banking access, and balance-sheet economics. Wallet infrastructure is increasingly competitive. Blockchain transaction fees tend toward commodity pricing. ChainIT\textquotesingle{}s durable position is the layer that binds the chosen monetary instrument to verified commercial context.

The highest-value system determines who may transact, for whom, under what authority, subject to which current evidence and policy, on what exact instruction, through which qualified wallet and rail, after satisfaction of which conditions, with what finality, and with what resulting evidence.

\subsection{Agnostic by architecture, selective by policy}
ChainIT is designed to remain issuer, chain, wallet, custody, and rail-agnostic. Agnostic does not mean every provider is acceptable. It means the commercial transaction is not structurally dependent on one provider and can move among qualified alternatives without losing identity, authority, compliance, evidence, or reconciliation.

\begin{itemize}
  \item \textbf{Settlement-asset independence.} The obligation remains denominated and governed independently from the selected stablecoin, tokenized deposit, or fiat rail.
  \item \textbf{Network independence.} Chain-specific fields and finality rules are isolated in controlled profiles rather than embedded throughout the commercial logic.
  \item \textbf{Custody independence.} The same authority and compliance decision may govern self-custody, ChainIT MPC, a qualified custodian, smart account, or bank wallet when the execution boundary enforces the required controls.
  \item \textbf{Provider portability.} Issuers, custodians, analytics providers, liquidity providers, node providers, banks, and rails may change without rewriting the transaction\textquotesingle{}s identity, legal, or evidentiary layer.
\end{itemize}
\vspace{-1.35bp}%
\subsection{The commercial value resides above and across the rail}
\href{https://chainit.com/product/digital-pay/}{ChainIT Pay} can monetize the services that make digital value commercially operable: individual and organizational verification; authority creation and query; runtime compliance evidence; wallet policy; secure custody integration; route and quote management; conversion and payout coordination through authorized partners; conditional settlement; \href{https://www.pactvera.com/}{Pactvera} execution; \href{https://chainit.com/}{Valitorum} evidence; ERP integration; exception handling; and merchant or marketplace operations.

The same architecture can support a stablecoin today, a tokenized deposit tomorrow, and a conventional bank rail whenever cost, risk, regulation, availability, or customer preference requires it.

\begin{figure}[H]
  \centering\vspace*{0.6bp}
  \includegraphics[width=363.57bp]{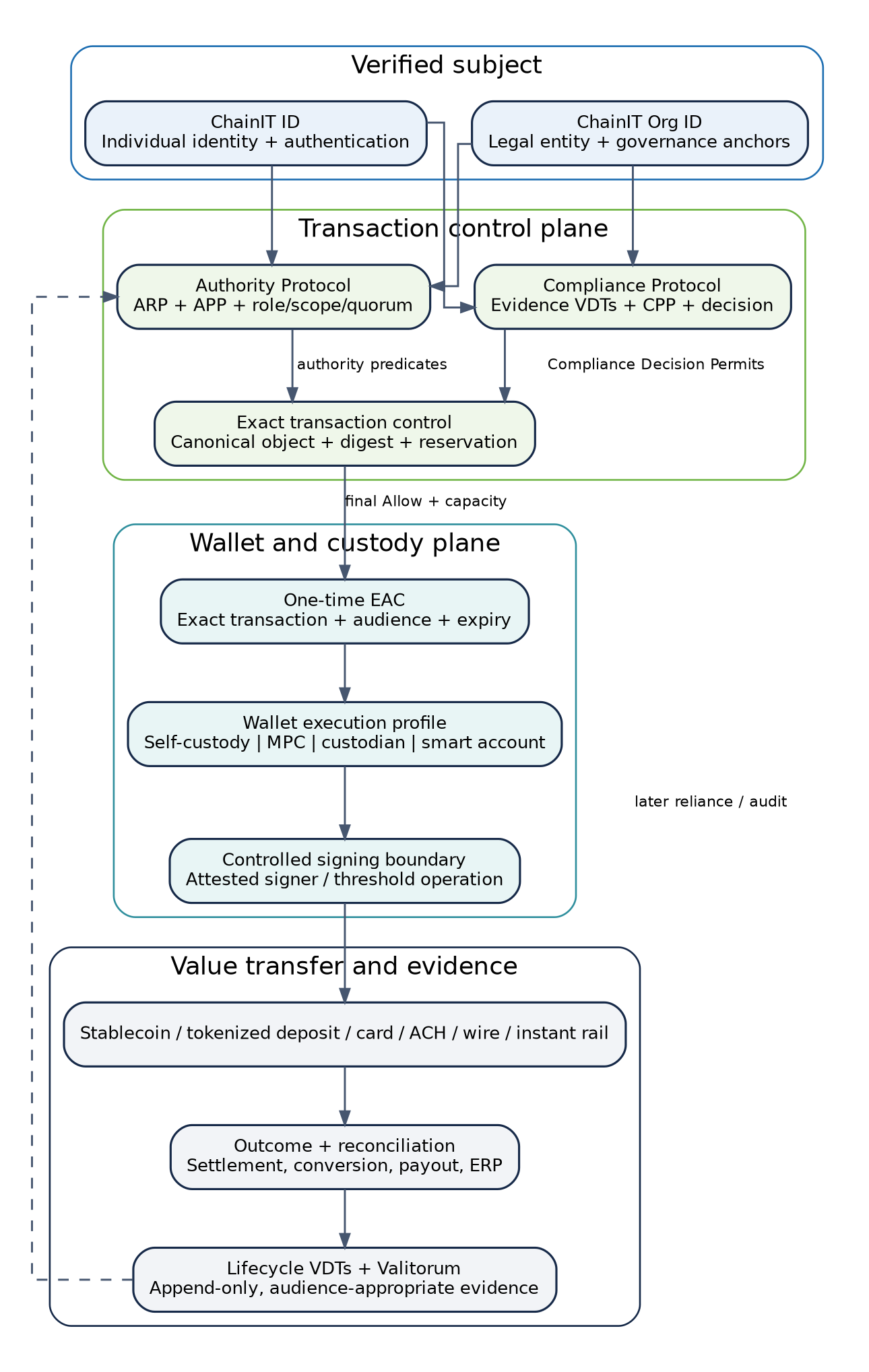}
  \caption{The ChainIT \href{https://chainit.com/white-papers/}{Transaction Truth} Architecture separates verified subjects, transaction control, wallet execution, value transfer, and lifecycle evidence.}\label{fig:1}\vspace{-5.5bp}
\end{figure}
\clearpage %
\section[How the Transaction Truth Architecture Operates]{\rule{0bp}{21.8bp}How the Transaction Truth Architecture Operates}%
\subsection{Four planes, one transaction}
\rowcolors{2}{white}{white}
\begin{longtable}{|@{\hspace{4bp}}P{203.8bp}|P{279.0bp}@{\hspace{4bp}}|}
\hline
\headrow\thc{Plane} & \thc{Executive function}\\\hline
\endfirsthead
\hline
\headrow\thc{Plane} & \thc{Executive function}\\\hline
\endhead
\textbf{Subject and transaction-control plane} & \href{https://chainit.com/product/identity-verification/}{ChainIT ID}, \href{https://chainit.com/product/business-verification/}{ChainIT Org ID}, ARP, APP, CPP, Evidence Assertion \href{https://chainit.com/product/data-tokenization/}{VDTs}, BRE predicates, Workflow Engine dispositions, approvals, and capacity reservation determine whether the exact action is eligible.\\\hline
\textbf{Wallet and custody plane} & The WEP, EAC validation, signer attestation, MPC threshold, smart-account policy, external custodian Application Programming Interface (API), or self-custodial signer determine whether a qualified wallet may produce the required authorization.\\\hline
\textbf{Value-transfer plane} & The blockchain, stablecoin issuer, tokenized-deposit platform, bank, card network, ACH operator, wire system, instant-payment rail, or smart contract moves or records value.\\\hline
\textbf{Evidence and integration plane} & \href{https://www.pactvera.com/}{Pactvera}, lifecycle \href{https://chainit.com/product/data-tokenization/}{VDTs}, \href{https://chainit.com/}{Valitorum}, \href{https://chainit.com/product/digital-pay/}{ChainIT Pay}, ERP connectors, treasury systems, and reporting preserve the decision and outcome in commercially meaningful form.\\\hline
\end{longtable}
\subsection{Execution is conjunctive}
A favorable result in one domain never substitutes for another. Strong MPC custody does not cure absent authority. A valid ARP does not cure a sanctions prohibition. A final compliance Allow does not cure a modified destination. A valid EAC does not cure an unapproved signer environment. A network confirmation does not cure failed merchant payout.

\ttformula{green}{Commit Eligible = Effective Authority AND Compliance Decision Permits AND Exact Digest Match AND Required Approvals AND Capacity Reserved AND Qualified Execution Boundary}
\subsection{Executive transaction sequence}
\begingroup\setlength{\padtop}{14.8bp}\setlength{\padbot}{7.2bp}%
\rowcolors{2}{white}{white}
\begin{longtable}{|@{\hspace{4bp}}P{491.3bp}@{\hspace{4bp}}|}
\hline
\headrow\thc{Control outcome}\\\hline
\endfirsthead
\hline
\headrow\thc{Control outcome}\\\hline
\endhead
Establish the verified individual, organization, group, workflow, or registered agent.\\\hline
Bind the relevant wallet, custody account, or payment source under a defined assurance method.\\\hline
Create the payment intent or \href{https://www.pactvera.com/}{Pactvera} transaction and identify the commercial obligation.\\\hline
Select current APP, CPP, and WEP versions and refresh required evidence.\\\hline
Resolve authority and compliance for this reliance event.\\\hline
Canonicalize the exact instruction and produce its digest.\\\hline
Complete required human, quorum, or delegated-agent approvals against the same digest.\\\hline
Reserve balance, budget, or aggregate capacity atomically.\\\hline
Issue one short-lived, audience-restricted, single-use EAC.\\\hline
Have the wallet boundary independently validate the EAC, digest, WEP, and signer state.\\\hline
Sign or authorize, broadcast through the selected rail, and monitor the outcome.\\\hline
Confirm conversion, payout, commercial completion, ERP reconciliation, and \href{https://chainit.com/}{Valitorum}.\\\hline
\end{longtable}
\endgroup
\subsection{Non-negotiable zero-trust invariants}
\begin{itemize}
  \item No Session Credential is payment authority.
  \item Every material transaction field is included in the canonical digest.
  \item Capacity is reserved before execution authority is issued.
  \item EACs are audience-restricted, short-lived, and single-use.
  \item The signing boundary independently reconstructs and verifies the digest (recomputes the fingerprint from the transaction it is about to sign and confirms it matches the one that was approved).
  \item Organizational approval quorum and cryptographic threshold are satisfied independently.
  \item Missing, stale, ambiguous, or contradictory evidence produces a controlled non-Allow state.
  \item Prior immutable records are never rewritten; later events are linked and appended.
  \item Sensitive evidence is disclosed only to the authorized audience and purpose.
  \item Automated systems may propose and coordinate but cannot bypass deterministic commit controls.
\end{itemize}
\vspace{-1.35bp}%
\section{Individual and Organizational Wallet Transactions}
\subsection{The control model is subject-neutral, but the authority source differs}
An individual may act for the individual\textquotesingle{}s own account or under a valid delegation. An organization can act only through persons, groups, scheduled workflows, or registered agents whose authority is derived from organizational governance. The transaction sequence is shared; the identity, authority, approval, and capacity evidence differs.

\rowcolors{2}{white}{rowalt}
\begin{longtable}{|@{\hspace{4bp}}P{139.5bp}|P{164.2bp}|P{170.6bp}@{\hspace{4bp}}|}
\hline
\headrow\thc{Control} & \thc{Individual transaction} & \thc{Organizational transaction}\\\hline
\endfirsthead
\hline
\headrow\thc{Control} & \thc{Individual transaction} & \thc{Organizational transaction}\\\hline
\endhead
\textbf{Principal} & The verified individual, unless acting for another principal. & The legal entity identified through \href{https://chainit.com/product/business-verification/}{ChainIT Org ID}.\\\hline
\textbf{Identity anchor} & \href{https://chainit.com/product/identity-verification/}{ChainIT ID} and wallet-control evidence. & \href{https://chainit.com/product/business-verification/}{ChainIT Org ID} plus ChainIT IDs for authorized representatives.\\\hline
\textbf{Authority source} & Personal control, legal capacity, or a valid delegation. & ARP, role, purpose, limits, quorum, separation of duties, and revocation.\\\hline
\textbf{Compliance context} & Individual, counterparty, wallet, asset, chain, jurisdiction, and transaction. & Organization, beneficial ownership, authorized persons, supplier or merchant, wallet, asset, chain, jurisdiction, and transaction.\\\hline
\textbf{Approval} & The individual approves the exact digest, with step-up when policy requires. & One or more authorized persons approve the same digest under the ARP quorum.\\\hline
\textbf{Capacity} & Available assets, personal limits, risk buffer, and active reservations. & Treasury balance, budget, entity and department limits, counterparty limits, aggregate exposure, and active reservations.\\\hline
\textbf{Execution} & Qualified self-custodial, MPC, custodial, smart-account, or bank wallet. & Qualified institutional, MPC, custodial, smart-account, bank, or embedded wallet under the WEP.\\\hline
\textbf{Evidence} & Identity, authority or delegation, decision, approval, execution, outcome, and \href{https://chainit.com/}{Valitorum}. & \href{https://chainit.com/product/business-verification/}{Org ID}, ARP, approvals, decision, reservation, signing, settlement, reconciliation, and Valitorum.\\\hline
\end{longtable}
\subsection{Organizational governance and cryptographic signing are orthogonal}
An organization may require two officers to approve a transfer while an MPC wallet uses a different threshold among infrastructure signers. The officers may not hold key shares. The cryptographic signers may not possess legal authority to approve the payment. The two layers answer different questions and must not be conflated.

\ttformula{green}{Authorized Organizational Transaction = Governance Quorum Satisfied AND Cryptographic Signing Threshold Satisfied}
\subsection{Agent and scheduled-workflow initiation}
A scheduled workflow or AI agent may create a proposal, retrieve evidence, calculate a route, or coordinate a transaction within a defined scope. It does not acquire authority merely because it can call a wallet API. A registered agent acts only under authority delegated by a verified principal, bounded by purpose, payee, transaction class, rail, amount, aggregate limit, time, step-up rules, and revocation. The exact transaction still passes through the same compliance, digest, reservation, EAC, wallet, and evidence controls. \cite{tt1}

Crossing the delegated boundary, (such as a new beneficiary, changed destination, increased value, irreversible rail, or contract interaction outside scope) requires a new decision and any prescribed human or quorum approval.

\section{Wallet and Custody Profiles}
\subsection{A wallet is a profile, not a single product category}
The term wallet describes materially different control systems. Some place a private key entirely under user control. Some delegate signing to a custodian. Some use multiple on-chain signatures. Some perform off-chain threshold cryptography and produce one conventional signature. Some are smart accounts whose policies execute in contract code. Others are interfaces to an exchange, bank, or internal ledger. ChainIT governs these systems through a WEP rather than assuming one wallet label implies one security model.

\rowcolors{2}{white}{rowalt}
\begin{longtable}{|@{\hspace{4bp}}P{138.3bp}|P{138.3bp}|P{197.7bp}@{\hspace{4bp}}|}
\hline
\headrow\thc{Wallet profile} & \thc{Primary control model} & \thc{ChainIT control emphasis}\\\hline
\endfirsthead
\hline
\headrow\thc{Wallet profile} & \thc{Primary control model} & \thc{ChainIT control emphasis}\\\hline
\endhead
\textbf{Individual self-custody} & User-controlled key or hardware signer. & Wallet binding, exact-digest presentation, beneficiary and contract controls, step-up approval, monitoring, and evidence of whether the transaction stayed inside the governed path.\\\hline
\textbf{ChainIT MPC profile} & Threshold signing using protected shares and approved signer domains. & EAC-validating coordinator and attested signer components; threshold, signer set, recovery, and evidence governed by the WEP.\\\hline
\textbf{External custodian} & Regulated or institutional provider controls keys and account policy. & Preflight authorization, signed sidecar or approval reference, account mapping, outcome reconciliation, and contractual responsibility allocation.\\\hline
\textbf{Smart account} & On-chain program evaluates execution conditions. & Approved policy module, contract and calldata controls, upgrade-key governance, EAC or equivalent preflight, and chain-specific evidence.\\\hline
\textbf{Embedded merchant or marketplace wallet} & Wallet provisioned within an application or commerce workflow. & \href{https://chainit.com/product/chainit-x/}{ChainIT X} or \href{https://chainit.com/product/digital-pay/}{ChainIT Pay} onboarding, transaction, custody boundary, merchant payout, recovery, and disclosure controls.\\\hline
\textbf{Bank or tokenized-deposit wallet} & Bank controls the account or tokenized deposit. & Bank-native authorization integration, entity eligibility, settlement terms, and reconciliation without misclassifying the instrument as a generic stablecoin.\\\hline
\end{longtable}
\subsection{MPC and Nitro Enclaves: custody security, not business authority}
MPC distributes signing capability so no single conventional private key or participant must act alone. The security benefit depends on the protocol, threshold, participant independence, distributed key generation, share protection, recovery, software integrity, and operating procedures. MPC reduces a class of single-key failures; it does not establish who is legally authorized to cause the signature.

The ChainIT MPC Wallet MVP profile uses Amazon Web Services (AWS) Nitro Enclaves as a confidential-computing boundary for approved signing logic and protected share operations. Enclaves can isolate memory from the parent instance, produce attestation documents identifying the measured software, and support policy-bound access to protected material through AWS Key Management Service (KMS). The wallet profile must still validate the EAC, exact digest, WEP, nonce, status, and attestation before any threshold participant signs. \cite{tt4,tt20}-[\citen{tt22}]

The detailed share count, threshold, cloud-account topology, region, key ceremony, recovery method, supported chains, and service levels remain controlled by the current internal engineering architecture and deployment configuration. The executive conclusion is simpler: protected signing components may execute only a transaction already approved by the independent authority and compliance control planes.

\clearpage %
\begin{figure}[H]
  \centering\vspace*{0.6bp}
  \includegraphics[width=449.89bp]{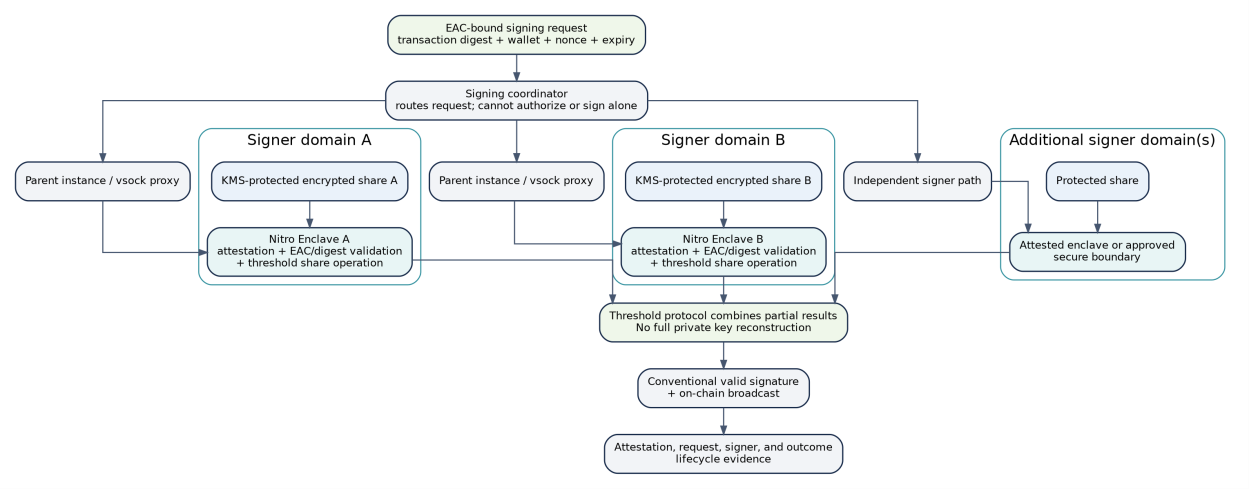}
  \caption{Abstract MPC profile: independent signer domains validate one approved transaction and contribute partial results without acquiring organizational authority.}\label{fig:2}\vspace{-5.5bp}
\end{figure}
\section{Runtime Compliance and Stablecoin Qualification}
\subsection{Compliance evidence remains attributable, current, and purpose-specific}
\href{https://chainit.com/white-papers/}{Provable Compliance} does not reduce a person or organization to a universal compliant or noncompliant label. Each sourced fact is represented as an Evidence Assertion \href{https://chainit.com/product/data-tokenization/}{VDT} with source class, provenance, observed time, effective time, freshness or expiration, jurisdiction, purpose, confidence or match state, and privacy classification. Evidence may come from an authoritative source, a qualified third party, or the institution itself. \cite{tt2}

The CPP selects the evidence and controls applicable to the institution, product, subject, counterparty, transaction type, value, rail, payer and payee jurisdiction, customer category, risk tier, and intended purpose. The BRE returns deterministic, reason-coded predicate results. The organization\textquotesingle{}s Workflow Engine composes those results into an explicit disposition such as Allow, Allow with Controls, Step-Up, Review, Hold, Reject, Prohibit, or Stale/Unknown. The wallet independently checks that the transaction in front of it is identical, detail for detail, to the one that was approved, and refuses to sign if anything has changed.

\subsection{Compliance does not confer authority}
\rowcolors{2}{white}{white}
\begin{longtable}{|@{\hspace{4bp}}P{241.4bp}|P{241.4bp}@{\hspace{4bp}}|}
\hline
\headrow\thc{Compliance Decision \href{https://chainit.com/product/data-tokenization/}{VDT} does} & \thc{Compliance Decision VDT does not}\\\hline
\endfirsthead
\hline
\headrow\thc{Compliance Decision \href{https://chainit.com/product/data-tokenization/}{VDT} does} & \thc{Compliance Decision VDT does not}\\\hline
\endhead
Records the issuing organization or authorized decision service, audience, CPP version, evidence references, freshness, reason codes, and disposition. & Create an ARP grant, change an officer\textquotesingle{}s role, expand a transaction limit, or establish governance authority.\\\hline
Determines whether current compliance-side evidence and policy permit continuation for one reliance event. & Authorize a wallet signature, consume capacity, select a final payment instruction, or bypass the \href{https://chainit.com/insights/chainit-launches-provable-authority-as-u-s-generative-ai-enabled-fraud-losses-are-projected-to-reach-40-billion-by-2027/}{Authority Protocol}.\\\hline
Supports controlled disclosure to the subject, institution, partner, auditor, regulator, or other authorized audience. & Turn restricted evidence into public blockchain data or disclose Suspicious Activity Related (SAR)-related information to an unauthorized party.\\\hline
Produces append-only review, override, and supersession evidence. & Rewrite an earlier evidence or decision record.\\\hline
\end{longtable}
\clearpage %
\addtocontents{toc}{\protect\newpage}%
\subsection{Stablecoin and Web3 transaction evidence domains}
\begin{itemize}
  \item Individual and organizational identity, beneficial ownership, entity status, and representative relationship;
  \item Authority, role, delegation, quorum, limits, and revocation;
  \item Customer and counterparty risk, sanctions, Anti-Money Laundering (AML)/Countering the Financing of Terrorism (CFT), transaction monitoring, and source-specific reviews;
  \item Wallet ownership or control, address and transaction risk, beneficiary history, and destination changes;
  \item Issuer authorization, reserve and redemption structure, token contract, freeze capability, and geographic eligibility;
  \item Backing-mechanism classification (off-chain collateralized, on-chain collateralized, uncollateralized/algorithmic, or hybrid/synthetic), reflecting materially different failure modes and control focus;
  \item Native, bridged, or wrapped asset status; chain security, finality, outages, and concentration;
  \item Custodian, bank, liquidity provider, conversion provider, and settlement pathway; and
  \item Transaction purpose, value, velocity, jurisdiction, product, rail, and policy-specific evidence.
\end{itemize}
\vspace{-4.8bp}%
A shared peg does not imply a shared risk profile. Collateralized and uncollateralized (algorithmic) stablecoins fail in different ways: an off-chain, fiat-collateralized instrument depends on issuer solvency and reserve attestation, an on-chain collateralized instrument depends on collateralization ratio and liquidation mechanics, and an algorithmic instrument depends on protocol incentives with no independent redeemable reserve. ChainIT records backing-mechanism classification as a first-class registry attribute, versioned like any other field, rather than inferring it from a stablecoin\textquotesingle{}s name or marketing description. Recent research reaches a consistent conclusion on why this matters: reserve quality and transparency, not the collateralization label alone, drive run risk, and overcollateralization plus disclosure requirements measurably reduce run and insolvency risk. \cite{tt24,tt25}

\subsection{Regulatory perimeter remains activity-specific}
The U.S. stablecoin framework reinforces the need to separate issuer obligations from transaction controls. The GENIUS Act establishes a federal framework for permitted payment stablecoin issuers, reserves, redemption, disclosure, supervision, AML/CFT, sanctions, and related requirements. Current rulemakings address issuer and customer-identification obligations, while Financial Crimes Enforcement Network (FinCEN), Office of Foreign Assets Control (OFAC), banking, custody, money-transmission, and institutional requirements continue to depend on the actual activity and flow of funds. \cite{tt11}-[\citen{tt18}]

ChainIT should maintain an explicit responsibility matrix identifying who receives funds, controls private keys, performs conversion, selects counterparties, holds merchant funds, executes sanctions and transaction monitoring, files reports, handles blocked property, manages refunds, bears settlement risk, and retains evidence. Regulated functions should be performed by the appropriately authorized ChainIT entity or qualified partner.

\subsection{Privacy and disclosure}
Compliance does not always require every merchant or counterparty to receive the underlying identity record. ChainIT may provide purpose-limited assertions such as identity verified, organization verified, authority current, sanctions screening completed, jurisdiction permitted, or transaction limit not exceeded, while preserving restricted evidence in the proper environment. The disclosure result must identify its audience, purpose, source, time, and limitations.

\clearpage %
\section{Exact-Transaction Execution}
\subsection{Approval must follow the transaction, not merely the session}
A user may be correctly authenticated while a transaction changes between presentation and signing. Destination substitution, fee modification, token-contract changes, altered calldata, changed chain, route fallback, quote expiration, or compromised middleware can cause the instruction executed by the wallet to differ from the instruction approved by the person or organization.

ChainIT addresses this through an exact-transaction control sequence rather than a general logged-in state.

\subsection{Canonicalization and digest}
The transaction object includes the payer, source wallet or account, payee, destination, amount, currency or asset, issuer, token contract, chain, method or calldata hash, nonce, fees, quote, validity window, commercial reference, and other material rail-specific fields. A fixed canonicalization method produces one digest. Any material change produces a different digest. \cite{tt23}

\subsection{Decision, approval, reservation, and EAC}
\rowcolors{2}{white}{rowalt}
\begin{longtable}{|@{\hspace{4bp}}P{205.2bp}|P{277.7bp}@{\hspace{4bp}}|}
\hline
\headrow\thc{Control} & \thc{Purpose}\\\hline
\endfirsthead
\hline
\headrow\thc{Control} & \thc{Purpose}\\\hline
\endhead
\textbf{Transaction Policy Decision VDT} & Binds the current authority and compliance result, risk tier, policy versions, evidence, conditions, and disposition to one exact digest.\\\hline
\textbf{Exact-digest approval} & The individual or every required organizational approver authorizes the same digest and material fields.\\\hline
\textbf{Atomic reservation} & Balance, budget, counterparty limit, aggregate exposure, or other capacity is reserved before execution authority is issued.\\\hline
\textbf{One-time EAC} & A short-lived credential authorizes one digest for one wallet or rail audience and cannot serve as a standing payment entitlement.\\\hline
\textbf{Wallet-bound verification} & The wallet or custodian independently reconstructs the transaction, validates the digest and EAC, and rejects mismatch, replay, wrong audience, stale status, or expiration.\\\hline
\textbf{Idempotent submission} & Retries do not create duplicate economic transfers, and an unknown result remains controlled until resolved.\\\hline
\end{longtable}
\subsection{Asset, chain, destination, and contract qualification}
The same exact-transaction decision must reference a qualified issuer, token contract, blockchain, custody profile, and destination. ChainIT should maintain versioned registries covering legal and operational status, native versus bridged or wrapped form, redemption path, freeze or blacklist capability, chain finality, bridge dependencies, supported custodians, conversion routes, liquidity, geographic restrictions, concentration limits, and emergency status.

Destination and smart-contract controls should distinguish an established verified beneficiary from a first-use address, a verified contract from an arbitrary call target, and a limited method from an unrestricted approval. A material change after approval invalidates the prior digest and requires a new decision. Fallback routing may change the rail only when the commercial terms and authorization permit that change; otherwise renewed authorization is required.

\begin{figure}[H]
  \centering\vspace*{0.6bp}
  \includegraphics[width=449.96bp]{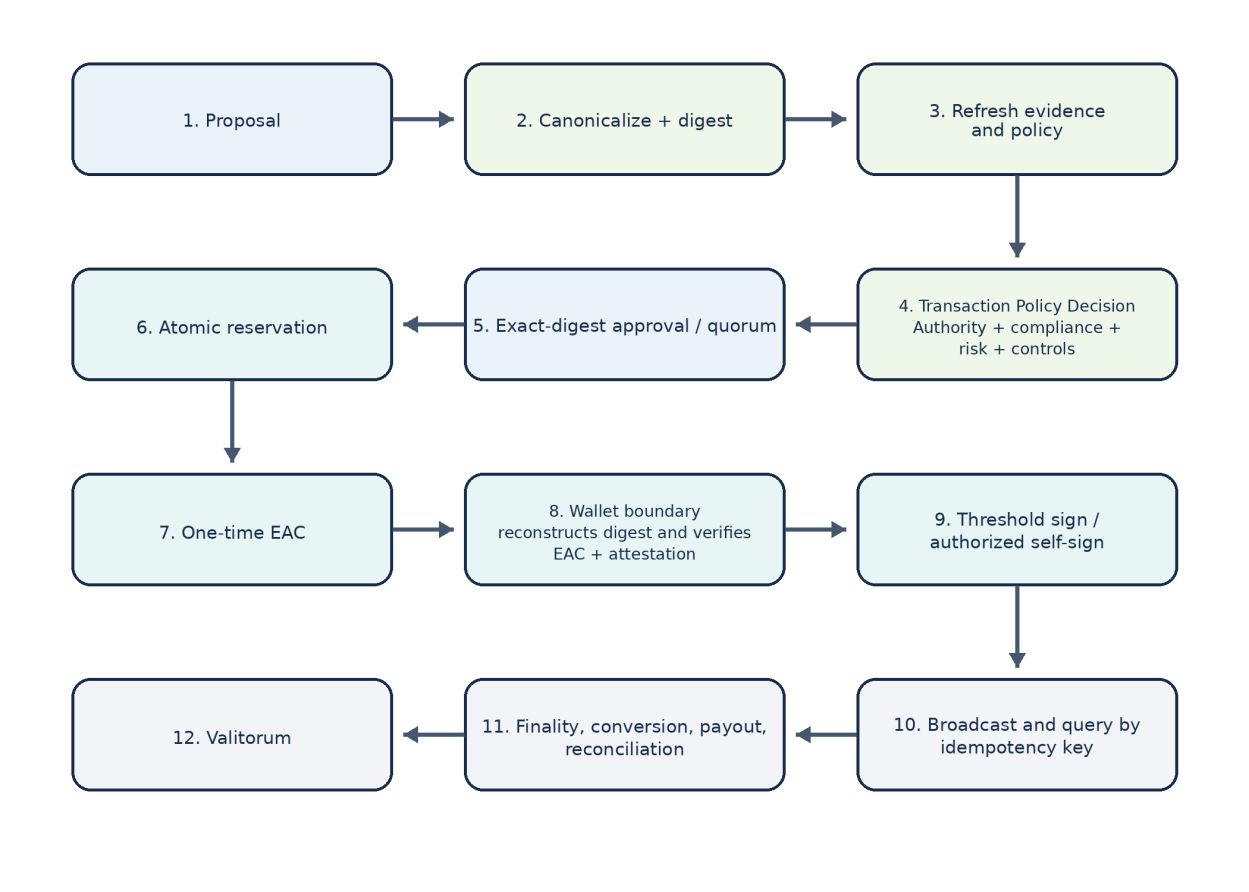}
  \caption{Exact-transaction commit sequence: proposal, current evidence and policy, decision, approval, reservation, one-time EAC, wallet validation, signing, settlement, reconciliation, and \href{https://chainit.com/}{Valitorum}.}\label{fig:3}\vspace{-5.5bp}
\end{figure}
\section{Settlement, Reconciliation, and Immutable Evidence}
\subsection{Finality occurs in layers}
\rowcolors{2}{white}{rowalt}
\begin{longtable}{|@{\hspace{4bp}}P{200.0bp}|P{282.8bp}@{\hspace{4bp}}|}
\hline
\headrow\thc{Finality layer} & \thc{Meaning}\\\hline
\endfirsthead
\hline
\headrow\thc{Finality layer} & \thc{Meaning}\\\hline
\endhead
\textbf{Signature finality} & The required key or threshold operation completed.\\\hline
\textbf{Network finality} & The transaction was accepted and reached the required consensus confidence.\\\hline
\textbf{Issuer finality} & The token remains recognized and transferable under issuer controls.\\\hline
\textbf{Conversion finality} & The digital asset was exchanged at the recorded rate and fees.\\\hline
\textbf{Bank or merchant payout} & The requested fiat or deposit proceeds were credited.\\\hline
\textbf{Commercial finality} & The underlying \href{https://www.pactvera.com/}{Pactvera} conditions, delivery, milestone, or other obligation were satisfied.\\\hline
\textbf{Accounting finality} & The transaction, fees, conversion, and payout were reconciled to the ERP and general ledger.\\\hline
\textbf{Evidentiary finality} & The required lifecycle evidence was sealed in \href{https://chainit.com/}{Valitorum}.\\\hline
\end{longtable}
A blockchain confirmation should not be presented as completed merchant settlement when conversion, payout, contractual performance, or reconciliation remains open.

\subsection{Pactvera, lifecycle VDTs, and Valitorum}
\href{https://www.pactvera.com/}{Pactvera} binds the transaction\textquotesingle{}s parties, authority, terms, consideration, conditions, payment intent, approvals, execution, and outcome. Lifecycle \href{https://chainit.com/product/data-tokenization/}{VDTs} record discrete sourced facts and state changes without rewriting prior entries. \href{https://chainit.com/}{Valitorum} seals the completed or otherwise terminal evidence chain under a defined completeness rule.

The evidence graph should answer who acted, for whom, under what authority, against which current evidence and policy, on what exact instruction, through which wallet and rail, what value moved, whether conversion and payout occurred, how the transaction was booked, and what later exception followed.

\subsection{Dual-ledger privacy and selective disclosure}
Sensitive commercial, identity, bank, tax, and compliance data may remain in protected storage while hashes, signatures, timestamps, state transitions, and integrity references are recorded in immutable or tamper-evident form. Authorized reviewers can verify the protected record without publishing restricted evidence to a public chain.

\subsection{Refunds and later exceptions}
An irreversible on-chain transfer is not erased by describing a later event as a reversal. A refund is modeled as a new compensating transaction linked to the original. The original \href{https://chainit.com/}{Valitorum} remains unchanged. A Supplemental Valitorum records the return, dispute, recall, recovery, chargeback, or other later exception and preserves the relationship between both economic events.

\begin{figure}[H]
  \centering\hspace*{-9.12bp}
  \includegraphics[width=234.57bp]{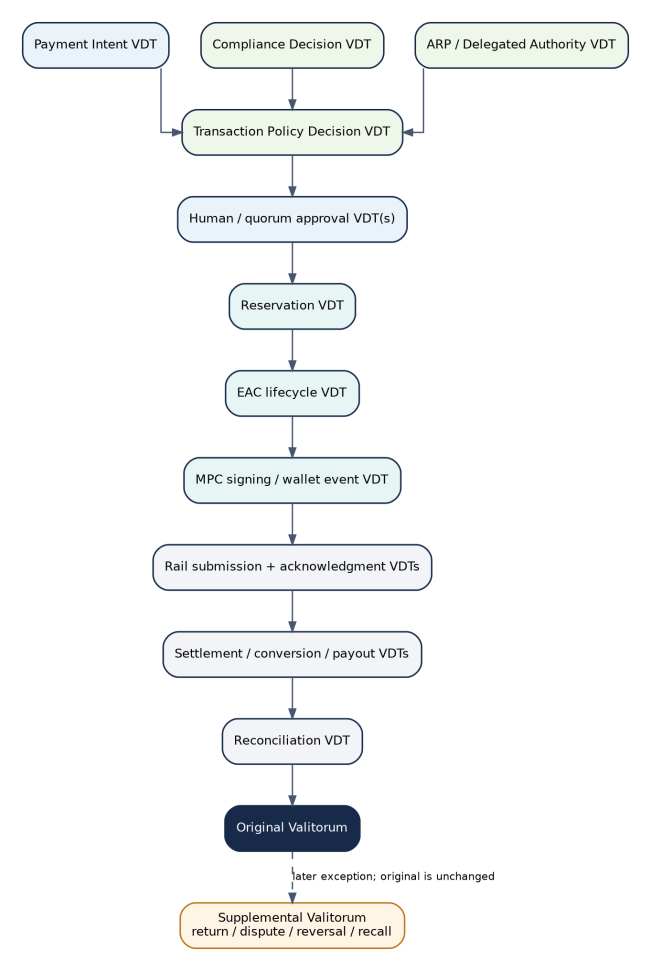}
  \caption{Append-only evidence graph: transaction intent, authority, compliance, approval, reservation, wallet execution, settlement, reconciliation, original \href{https://chainit.com/}{Valitorum}, and linked later exceptions.}\label{fig:4}\vspace{-5.5bp}
\end{figure}
\section{Illustrative Commercial Use Cases}
\vspace{-0.8bp}%
\rowcolors{2}{white}{rowalt}
\begin{longtable}{|@{\hspace{4bp}}P{140.2bp}|P{140.3bp}|P{193.9bp}@{\hspace{4bp}}|}
\hline
\headrow\thc{Use case} & \thc{Business objective} & \thc{ChainIT control and evidence}\\\hline
\endfirsthead
\hline
\headrow\thc{Use case} & \thc{Business objective} & \thc{ChainIT control and evidence}\\\hline
\endhead
\textbf{Merchant stablecoin acceptance with fiat settlement} & A customer pays an approved stablecoin; the merchant receives fiat. & Customer and merchant identity, asset and chain qualification, wallet screening, exact payment intent, conversion, merchant payout, ERP reconciliation, and \href{https://chainit.com/}{Valitorum}.\\\hline
\textbf{Organizational supplier payment through MPC} & A corporation pays an approved supplier under a two-approver treasury policy. & \href{https://chainit.com/product/business-verification/}{ChainIT Org IDs}, ARP roles, supplier evidence, same-digest quorum, atomic capacity, EAC, attested threshold signing, finality, invoice reconciliation, and evidence.\\\hline
\textbf{Conditional marketplace settlement} & Funds are held or reserved until asset, delivery, inspection, or acceptance conditions are satisfied. & Verified buyer and seller, \href{https://chainit.com/product/data-tokenization/}{VDT}-linked asset, \href{https://www.pactvera.com/}{Pactvera} conditions, compliance, escrow or conditional release, multi-party allocation, and Supplemental \href{https://chainit.com/}{Valitorum} for later disputes.\\\hline
\textbf{Tokenized-deposit treasury transfer} & Institutions move value through a bank-issued digital deposit instrument. & Instrument classification, bank eligibility, \href{https://chainit.com/product/business-verification/}{Org ID}, authority, transaction policy, bank-native execution, treasury reconciliation, and evidence without treating the deposit as a generic stablecoin.\\\hline
\textbf{Agent-initiated commerce} & A registered agent proposes a purchase or payment within delegated bounds. & Agent Registration \href{https://chainit.com/product/data-tokenization/}{VDT}, Agent Authority VDT, sponsoring organization, purpose and budget limits, current compliance, exact digest, step-up outside scope, EAC, and sovereign wallet boundary.\\\hline
\textbf{Cross-border B2B settlement} & A buyer settles an invoice through a qualified digital or bank rail selected for cost, time, jurisdiction, and risk. & Counterparty and beneficial-owner evidence, route and quote, sanctions and payment transparency, exact instruction, conversion or local payout, ERP posting, and evidence.\\\hline
\end{longtable}
\subsection{What changes for the customer}
The merchant or enterprise does not have to treat every wallet, stablecoin, chain, or provider as a separate operating model. ChainIT presents one controlled transaction lifecycle: verified parties, current authority, transaction-specific compliance, qualified settlement options, exact approval, secure signing, normalized status, reconciliation, and auditable evidence.

The customer may select a preferred custody model or settlement asset, while the \href{https://chainit.com/white-papers/}{Transaction Truth} Architecture preserves the same governance and evidentiary requirements across qualified alternatives.

\section{Business Value, Competitive Position, and IP Foundation}
\subsection{From wallet utility to enterprise control surface}
Conventional wallet products compete on custody, asset support, user experience, and transaction fees. ChainIT adds the enterprise control surface: verified individual and organizational identity; portable authority; runtime compliance; exact transaction approval; secure wallet execution; multi-rail orchestration; conditional commercial logic; ERP reconciliation; and immutable evidence.

The value proposition changes from \textquotedbl{}store and send tokens\textquotedbl{} to \textquotedbl{}execute a verified commercial transaction under provable authority and compliance.\textquotedbl{}

\subsection{Revenue across the transaction stack}
\rowcolors{2}{white}{white}
\begin{longtable}{|@{\hspace{4bp}}P{203.8bp}|P{279.0bp}@{\hspace{4bp}}|}
\hline
\headrow\thc{Revenue layer} & \thc{Representative services}\\\hline
\endfirsthead
\hline
\headrow\thc{Revenue layer} & \thc{Representative services}\\\hline
\endhead
\textbf{Identity and organization} & \href{https://chainit.com/product/identity-verification/}{ChainIT ID}, \href{https://chainit.com/product/business-verification/}{ChainIT Org ID}, authentication, Know Your Business (KYB), beneficial-owner, and attribute-verification services.\\\hline
\textbf{Authority and compliance} & ARP creation and query, delegation, revocation, evidence ingestion, policy execution, decisioning, and controlled disclosure.\\\hline
\textbf{Wallet and custody} & Wallet provisioning, MPC profile, custody integration, policy administration, recovery, and transaction control.\\\hline
\textbf{Payment and settlement} & Multi-rail orchestration, stablecoin acceptance, route and quote services, conversion and payout coordination, and treasury operations through authorized providers.\\\hline
\textbf{Transaction logic and evidence} & \href{https://www.pactvera.com/}{Pactvera}, conditional settlement, \href{https://chainit.com/product/data-tokenization/}{VDT} lifecycle records, \href{https://chainit.com/}{Valitorum} generation and verification, disputes, and Supplemental Valitorum.\\\hline
\textbf{Enterprise integration} & ERP, treasury, marketplace, reporting, monitoring, analytics, and exception management.\\\hline
\end{longtable}
\subsection{Competitive differentiation}
\rowcolors{2}{white}{rowalt}
\begin{longtable}{|@{\hspace{4bp}}P{143.1bp}|P{143.1bp}|P{188.1bp}@{\hspace{4bp}}|}
\hline
\headrow\thc{Provider category} & \thc{Commonly controls} & \thc{ChainIT adds}\\\hline
\endfirsthead
\hline
\headrow\thc{Provider category} & \thc{Commonly controls} & \thc{ChainIT adds}\\\hline
\endhead
\textbf{Stablecoin issuer} & Issuance, reserves, redemption, and token controls. & Verified parties, authority, policy, wallet governance, multi-rail choice, reconciliation, and evidence.\\\hline
\textbf{Wallet or custodian} & Keys, signing, accounts, and asset storage. & Legal and organizational context, current compliance, exact-digest authority, \href{https://www.pactvera.com/}{Pactvera}, and \href{https://chainit.com/}{Valitorum}.\\\hline
\textbf{Blockchain} & Consensus and state transition. & Commercial meaning, off-chain evidence, authority, policy, payout, and ERP outcome.\\\hline
\textbf{Identity or compliance provider} & Identity proofing, screening, or monitoring result. & Portable organizational authority, transaction-specific composition, commit-time enforcement, settlement, and evidence.\\\hline
\textbf{Payment processor} & Authorization and settlement on supported rails. & Web3 wallet control, stablecoin and tokenized-value support, programmable conditions, and immutable lifecycle evidence.\\\hline
\textbf{ERP or treasury system} & Accounting, approval, cash management, and reporting. & Verified subject, portable authority, wallet and chain execution, and cryptographic outcome evidence.\\\hline
\end{longtable}
\subsection{Intellectual-property foundation}
The ChainIT \href{https://chainit.com/white-papers/}{Transaction Truth} Architecture builds on the documented ChainIT portfolio and applications addressing verified individuals and events; portable privacy-preserving identity; immutable representations and \href{https://chainit.com/product/data-tokenization/}{VDTs}; paired physical and digital assets; digital financial events and multiple payment mechanisms; smart-contract conditions and payment release; \href{https://www.pactvera.com/}{Pactvera} transaction formation and conditional compensation; marketplace and wallet integration; authority, revocation, and compliance; and \href{https://chainit.com/}{Valitorum} evidence. \cite{tt5}-[\citen{tt10}]

U.S. Patent No. 12,321,940 describes verification and immutable association of financial events and payment records, with specifications encompassing stablecoin, cryptocurrency, ACH, Electronic Funds Transfer (EFT), card, bank-transfer, and wallet-transfer mechanisms. The Marketplace and \href{https://www.pactvera.com/}{Pactvera} materials provide additional foundations for verified identity, token and value transfer, conditional logic, payment routing, authority, compliance, and immutable transaction evidence. \cite{tt6}-[\citen{tt9}]

References to patents and applications identify technical source foundations. They do not represent that every implementation described here is covered by an issued claim or currently deployed. Wallet-specific and integrated protocol mechanisms remain subject to patent-counsel review for written description, enablement, continuity, priority, inventorship, and filing route.

\section{Controlled Deployment and Governance}
\subsection{Architecture status}
\rowcolors{2}{white}{rowalt}
\begin{longtable}{|@{\hspace{4bp}}P{201.2bp}|P{281.6bp}@{\hspace{4bp}}|}
\hline
\headrow\thc{Status class} & \thc{Public meaning}\\\hline
\endfirsthead
\hline
\headrow\thc{Status class} & \thc{Public meaning}\\\hline
\endhead
\textbf{Current platform foundation} & \href{https://chainit.com/product/identity-verification/}{ChainIT ID}, \href{https://chainit.com/product/business-verification/}{ChainIT Org ID}, \href{https://chainit.com/product/digital-pay/}{ChainIT Pay}, \href{https://www.pactvera.com/}{Pactvera}, \href{https://chainit.com/product/data-tokenization/}{VDTs}, \href{https://chainit.com/}{Valitorum}, and related documented platform capabilities used as the foundation.\\\hline
\textbf{Published protocol architecture} & \href{https://chainit.com/insights/chainit-launches-provable-authority-as-u-s-generative-ai-enabled-fraud-losses-are-projected-to-reach-40-billion-by-2027/}{Provable Authority} and \href{https://chainit.com/white-papers/}{Provable Compliance} definitions, evidence, policy, decision, reservation, EAC, and lifecycle controls incorporated by reference.\\\hline
\textbf{Development / MVP} & MPC wallet components, integrations, and controlled validation work not represented as generally available in every environment.\\\hline
\textbf{Proposed execution profile} & WEP, signer-domain EAC enforcement, attestation-bound share access, asset and chain registries, and other wallet-specific controls defined for implementation and testing.\\\hline
\textbf{External dependency} & Stablecoin issuers, blockchains, banks, custodians, cloud providers, analytics providers, liquidity providers, and other regulated or technical partners.\\\hline
\end{longtable}
\subsection{Phased deployment}
\rowcolors{2}{white}{rowalt}
\begin{longtable}{|@{\hspace{4bp}}P{196.3bp}|P{286.5bp}@{\hspace{4bp}}|}
\hline
\headrow\thc{Phase} & \thc{Required outcome}\\\hline
\endfirsthead
\hline
\headrow\thc{Phase} & \thc{Required outcome}\\\hline
\endhead
\textbf{0 - Responsibility and architecture clearance} & Finalize entity scope, flow of funds, control of keys, regulatory responsibility, Authority-Compliance interface, canonical transaction schema, WEP, registries, evidence model, incident controls, and patent/public-disclosure clearance.\\\hline
\textbf{1 - MPC component validation} & Prove attestation, KMS conditions, share protection, threshold signing, coordinator non-sovereignty, digest and EAC validation, replay rejection, outage behavior, rotation, recovery, and emergency suspension without live customer funds.\\\hline
\textbf{2 - Individual wallet pilot} & Deploy a bounded individual profile with approved assets and chains, identity and wallet binding, current compliance, exact-digest approval, controlled signing, monitoring, reconciliation, and \href{https://chainit.com/}{Valitorum}.\\\hline
\textbf{3 - Organizational wallet pilot} & Add \href{https://chainit.com/product/business-verification/}{ChainIT Org ID}, ARP roles, maker-checker or quorum approval, aggregate capacity, MPC or custodian execution, supplier controls, ERP integration, revocation, and recovery.\\\hline
\textbf{4 - Stablecoin acceptance and conversion} & Add merchant acceptance, route and quote, stablecoin receipt, authorized conversion, fiat payout, refunds, exception handling, and settlement evidence.\\\hline
\textbf{5 - Multi-rail and institutional interoperability} & Expand to cards, ACH, wires, instant payments, tokenized deposits, additional qualified wallets, banks, custodians, and treasury systems.\\\hline
\textbf{6 - Controlled agentic execution} & Permit registered agents to propose and coordinate within delegated scope while retaining deterministic authority, compliance, capacity, exact-transaction, wallet, and evidence controls.\\\hline
\end{longtable}
\subsection{Pilot success means invariants hold under attack and failure}
A successful pilot is not a favorable demonstration. It proves that required invariants remain true under transaction mutation, replay, stale evidence, revoked authority, concurrent requests, wrong audience, signer outage, unapproved enclave measurement, destination substitution, provider failure, policy change, and disclosure-boundary probing.

\begin{itemize}
  \item Every governed material-field mutation changes the digest and is rejected by the execution boundary.
  \item A consumed, expired, canceled, wrong-audience, wrong-wallet, or replayed EAC cannot authorize signing.
  \item A session or ordinary authenticated state cannot trigger payment.
  \item Revoked, expired, out-of-scope, over-limit, wrong-payee, wrong-rail, or insufficient-quorum transactions remain non-executable.
  \item Parallel requests cannot consume the same capacity or exceed aggregate limits.
  \item Unapproved signer software or stale attestation cannot access protected signing material.
  \item Settlement, conversion, merchant payout, ERP reconciliation, and \href{https://chainit.com/}{Valitorum} can be traced to the same transaction.
  \item Restricted evidence remains within its authorized audience and purpose.
\end{itemize}
\vspace{-4.8bp}%
Executive governance should include Product, Engineering, Payments, Information Security, Compliance, Legal, Finance, Treasury, Risk, Operations, and senior management. New assets, issuers, chains, custodians, liquidity providers, risk limits, finality policies, emergency actions, material BRE or Workflow changes, and production deployment claims should require explicit approval and recorded ownership.

\section{Related Work}\label{sec:related}

\textbf{Threshold signing and custody.} Threshold ECDSA protocols let several parties jointly produce one signature without assembling the private key~\cite{tt27,tt28}, and NIST's multi-party threshold cryptography project is evaluating such schemes for standardization~\cite{tt20}. The MPC profile in Section~6.2 relies on this property for custody security. This paper keeps it separate from business authority: a threshold signature shows that enough signer shares cooperated, not that the organization authorized the transaction (Section~5.2).

\textbf{Smart accounts and agent payments.} Account abstraction (ERC-4337) lets a smart-contract wallet apply programmable validation rules before execution~\cite{tt29}, and agent-payment protocols such as AP2, ACP, and x402 standardize how software agents present payment mandates, check out, or pay for resources~\cite{tt30,tt31,tt32}. These mechanisms operate at the wallet or protocol layer; the \href{https://chainit.com/white-papers/}{Transaction Truth} Architecture places verified identity, organizational authority, current compliance, and exact-digest approval in front of them (Sections~4 and~8).

\textbf{Stablecoin stability and regulation.} Economic and legal work studies what keeps stablecoins stable and how their run risk can be regulated~\cite{tt33,tt34}, alongside the studies cited in Section~7.3~\cite{tt25,tt26}. These analyses concern the instrument. This paper treats a stablecoin as one qualified settlement asset among several and records issuer and backing characteristics as transaction evidence (Section~7.3).

\textbf{Transaction binding and attestation.} Dynamic linking under the EU strong-customer-authentication rules binds an authentication code to the amount and payee of a payment~\cite{tt35}, and remote-attestation architectures let a relying party appraise the state of a signing environment~\cite{tt36}. Exact-digest approval (Section~8.2) and attestation-bound share access (Sections~6.2 and~12.1) apply these ideas to wallet execution. The companion papers specify the authority and compliance layers used here~\cite{tt1,tt2}.

\section{Conclusion}
Stablecoins and Web3 wallets improve the mechanics of moving value. They do not eliminate the need to establish who is acting, for whom, under what authority, against which current evidence, subject to what policy, on what exact transaction, through what custody system, with what commercial outcome, and with what proof.

For an individual, ChainIT binds verified identity, wallet control, personal or delegated authority, current compliance, exact-digest approval, qualified signing, settlement, and evidence.

For an organization, ChainIT binds \href{https://chainit.com/product/business-verification/}{ChainIT Org ID}, the ARP, delegated roles, quorum, separation of duties, capacity, current compliance, the exact transaction, MPC or custodian policy, authenticated settlement, ERP reconciliation, and \href{https://chainit.com/}{Valitorum}.

MPC and confidential computing strengthen custody by distributing signing capability and isolating sensitive operations. They do not create authority. The \href{https://chainit.com/insights/chainit-launches-provable-authority-as-u-s-generative-ai-enabled-fraud-losses-are-projected-to-reach-40-billion-by-2027/}{Authority Protocol} determines who may act. The Compliance Protocol determines whether current evidence and policy permit continuation. Exact-transaction controls determine what instruction is approved. The WEP determines how a qualified wallet may sign. The rail moves value. \href{https://chainit.com/}{Valitorum} proves the resulting lifecycle.

\ttbox{green}{ChainIT does not need to become the stablecoin, the blockchain, or the custodian. It must become the control and evidence infrastructure through which qualified money and wallets become commercially verifiable.}
That position allows ChainIT to remain issuer-agnostic, chain-agnostic, wallet-agnostic, bank-compatible, tokenized-deposit compatible, and multi-rail while preserving one consistent model for verified commerce.

Stablecoins move value. Wallets sign instructions. ChainIT establishes who may act, whether the action is permitted now, what exact transaction may be executed, how it settles, and what proves the result.

\clearpage %
\section*{\rule{0bp}{22.3bp}References and Source Basis}\addcontentsline{toc}{section}{References and Source Basis}%
\begingroup
\refsize

\endgroup

\clearpage
\section*{\rule{0bp}{22.3bp}Important Notices}\addcontentsline{toc}{section}{Important Notices}%

This document is an executive strategic and technical white paper. It is not legal, regulatory, banking, custody, accounting, tax, investment, security-certification, or patentability advice. It does not replace a regulated institution\textquotesingle{}s AML/CFT, sanctions, customer-identification, custody, cybersecurity, or risk-management program.

The legal characterization of a stablecoin, tokenized deposit, wallet, custody arrangement, conversion service, smart account, payment activity, escrow structure, or transaction-orchestration service depends on applicable law and the specific facts, contracts, flow of funds, control of assets, control of private keys, and activities performed.

References to current or proposed U.S. stablecoin requirements reflect sources identified in this paper as of September 2026. Proposed rules may change before adoption. ChainIT, its customers, and its partners must evaluate final rules and jurisdiction-specific requirements before deployment.

References to ChainIT patents and patent applications describe issued claims, disclosed embodiments, pending claims, or technical source foundations. They do not represent that every described feature is currently deployed or that every commercial implementation falls within an issued claim.

References to immutable records mean records designed to be tamper-evident, chronologically verifiable, and resistant to unauthorized alteration. Immutability does not establish that original source data was accurate unless source, provenance, capture method, and verification were themselves validated.

References to MPC, threshold cryptography, or confidential computing do not imply that every implementation provides identical security. Security depends on the protocol, threshold, participant count, key-generation process, implementation, adversary model, signer independence, recovery process, attestation policy, and operational controls.

\href{https://www.pactvera.com/}{Pactvera}, \href{https://chainit.com/product/data-tokenization/}{VDTs}, and \href{https://chainit.com/}{Valitorum} are designed to capture and preserve evidence relevant to identity, authority, assent, consideration, conditions, execution, and outcome. Actual enforceability and admissibility remain subject to applicable law, jurisdiction, transaction structure, and judicial or administrative determination.

Stablecoins should not automatically be characterized as legal tender, insured deposits, cash, cash equivalents, risk-free assets, or obligations of the United States. Issuer, reserve, redemption, liquidity, depeg, freeze, custody, chain, bridge, smart-contract, sanctions, and operational risks must be evaluated separately. Depeg risk is not limited to algorithmic designs: market-based estimates for one major fiat-collateralized issuer put the average annual probability of a full depeg at roughly 60 basis points, rising above 200 basis points during the 2022 Terra-Luna stress episode, with market volatility and elevated interest rates as leading indicators. [26]

ChainIT should perform regulated financial functions directly only where appropriately authorized and should otherwise integrate properly licensed or regulated issuers, banks, custodians, acquirers, money transmitters, liquidity providers, and other partners.

\end{document}